# Microwave Power, DC Magnetic Field, Frequency and Temperature Dependence of the Surface Resistance of MgB$_2$

A. A. Zhukov[†], A. Purnell, Y. Miyoshi, Y. Bugoslavsky[‡], Z. Lockman, A. Berenov,

J. L. MacManus-Driscoll and L. F. Cohen

*Center for High Temperature Superconductivity, Imperial College of Science Technology*

*and Medicine, Prince Consort Road, London SW7 2BZ, U.K.*

[†] *Institute of Solid State Physics, Russian Academy of Science, Chernogolovka, 142432*

*Russia*

[‡] *General Physics Institute, Moscow, Russia*

H. Y. Zhai, H. M. Christen, M. P. Paranthaman and D. H. Lowndes

*Oak Ridge National Laboratory, Oak Ridge, TN 37931-6056*

M. H. Jo and M. G. Blamire

*Department of Materials Science, Cambridge University,*

*Cambridge CB2 3QZ*

Ling Hao and J. C. Gallop

*National Physical Laboratory, Queens Rd, Teddington, U.K.*

## Abstract

The microwave power, dc magnetic field, frequency and temperature dependence of the surface resistance of MgB$_2$ films and powder samples were studied. Sample quality is relatively easy to identify by a number of characteristics, the most clear being the breakdown in the $\omega^2$ law for poor quality samples. Analysis of the experimental data suggests the most attractive procedure for high quality film growth for technical applications.

# Introduction

The discovery of superconductivity in $MgB_2$ [1] has stimulated interest in the possibility of commercial use of this compound in its superconducting state. A number of papers have already been dedicated to obtaining high critical current density ($J_c$) on different kinds of samples, such as powder [2], powder in tube [3, 4], tapes [5] and thin films [6]. The material looks rather promising for niche applications because grain boundaries do not appear to be deleterious to the continuity of the current path (in contrast to high temperature superconductors). Further work on Josephson junction properties [7, 8] suggest these materials may also be promising candidates for a range of electronic applications.

Microwave surface impedance measurements are extremely interesting because the order parameter in $MgB_2$ is thought to be of s-wave symmetry which suggests that the residual losses ought to be low and the temperature dependence of the surface resistance ($R_s(T)$) ought to be weak, similar to the classical superconductors such as Nb [9], but usable at a 25 - 30 K operating temperature. Most of the microwave data on $MgB_2$, reports the presence of the small gap in the energy excitation spectrum. This conclusion comes from the strong temperature dependence of the field penetration depth $\lambda(T)$ and surface resistance measured on powder [10, 11], wire [12], pellets [13] and film [14]. However, on high quality partially c axis oriented films, the strong coupling regime in $\lambda(T)$ and the saturation of the $R_s(T)$ were observed [15].

In this paper $R_s(T)$ has been measured on different $MgB_2$ samples at different frequencies using dielectric puck resonator and parallel plate techniques. In case of the low quality samples (high residual surface resistance ($R_{res} = R_s(T = 0)$)) with steep slopes of $R_s(T)$ and $\lambda(T)$, the $\omega^2$ scaling law breaks down at high temperatures, whilst for high quality films that show strong coupling behavior of $\lambda(T)$, $R_s(T)$ satisfies the frequency scaling law over a wide temperature range.

**Experimental Details**

**Samples**

Samples used for the measurements are the powder sample (see [10] for details) and two films, Film I (random orientated, $T_c$ = 30 K) and Film II (partially c-axis oriented, $T_c$ = 38 K).

Film I (4 mm × 6 mm × 300 nm) was cold-grown on a c-plane sapphire substrate by pulsed laser deposition (PLD) with post annealing in Ar+$H_2$ atmosphere at 750°C. Oxygen partial pressure was kept low by using reducing gas mixtures, and by using Zr foil-gettering of oxygen. Both Mg powder and foil were also used in the vicinity of the film. For additional details see [16], CAM6 film.

Film II (4.5 mm × 10 mm × 750 nm) was prepared by e-beam evaporation of boron on sapphire r-plane and then it was *ex-situ* annealed in Mg vapor for 1 hour at 900°C. For additional details of film growth see [17] and reference therein, film type B.

**Experimental set up**

Experiments were performed with dielectric puck and parallel plate resonator techniques. Three dielectric puck resonators were used at three frequencies 2.5 ($TiO_2$), 3.5 ($TiO_2$) and 8.7 GHz (Alumina) inserted into a copper housing. The fundamental $TE_{011}$ mode with magnetic field aligned along the axis of cylinder was used for all measurements. The copper housing was attached to a closed cycle cooler with temperature range of operation 10 - 300K. Further details of the dielectric puck rig are given elsewhere [18].

The powder sample was attached directly to the dielectric puck with optical clean liquid (Opticlean). To measure the films an additional quartz spacer was used to decrease the sensitivity of the puck artificially. Films were attached to the spacer with Apiezon N vacuum grease.

Measuring the temperature dependence of the quality factor and frequency shift of the empty resonator ($Q_o(T)$ and $\Delta f_o(T)$) and of the resonator with the sample ($Q_s(T)$ and $\Delta f_s(T)$) we can extract the $R_s(T)$ and $\Delta\lambda(T)$ using the relation

$$R_s + i\omega\mu_0\Delta\lambda(T) = G_s\left[\frac{1}{Q(T)} - \frac{1}{Q_o(T)} + i\frac{(f_o(T)-f(T))}{2f}\right] \quad (1)$$

where $G_s$ is the geometry factor. Calculations to determine $G_s$ were performed with MAFIA (commercial available software) [19].

Film II was measured with parallel-plate techniques at 5.7 GHz. The second film was a commercial YBCO film (10 mm × 5 mm × 500 nm) produced by Du Pont ($R_s^{YBCO}$(21 K, 5GHz) = 20μΩ, $R_s^{YBCO}$(50 K, 5GHz) = 40μΩ). There are three different types of losses in parallel plate techniques: losses in the film surface ($1/Q_s(T)$), radiation losses ($1/Q_{rad}$), and losses in dielectric spacer ($1/Q_{tg(\delta)}$). Knowledge of the $R_s^{YBCO}(T)$, $1/Q_{rad}$, $1/Q_{tg(\delta)}$ and $G_s$ allows to extract $R_s(T)$ of the sample under investigation from $Q(T)$ data:

$$\Delta R_s(T) = G_s\left(\frac{1}{Q(T)} - \frac{1}{Q_{rad}} - \frac{1}{Q_{tg(\delta)}}\right) \quad (2)$$

DC resistivity measurements were performed on the films using a standard four point technique.

**Results**

Figure 1 shows the dc resistivity of films I and II. Film I shows a residual resistivity ratio (RRR) = 1.1 with $\rho(300\ K) = 293\ \mu\Omega\cdot cm$ and $\rho(T_c) = 264\ \mu\Omega\cdot cm$. The superconducting transition of Film I is broad $\delta T_c = 5\ K$. Film II looks more like good metal with RRR = 1.8, $\rho(300\ K) = 34\ \mu\Omega\cdot cm$ and $\rho(T_c) = 18.8\ \mu\Omega\cdot cm$. Film II has much sharper superconducting transition, $\delta T_c = 0.3\ K$. Using the formula for the normal skin depth $\rho(T_c) = R_n(T_c)^2 / \pi\mu_o f_o$ where $f_o$ is the resonance frequency and $\mu_o$ is the permeability, then for the powder sample $\rho(T_c) = 25\ \mu\Omega\cdot cm$ [10]. The powder sample and Film I are in dirty limit, Film II is in the intermediate limit [15].

Figure 2 shows the $R_s(T)$ for all the samples. All data are rescaled to 10GHz using the $\omega^2$ law. Low quality samples such as the powder and Film I have high losses with typical values of $R_s(10\ K, 10\ GHz) \sim 5\ m\Omega$ at 10 K. For the higher quality Film II $R_s(10\ K, 10\ GHz) \sim 1\ m\Omega$. In general, there are two features of the $R_s(T,\omega)$ that can be distinguished. The first is the steeper slopes of $R_s(T)$ of the poor quality samples in contrast to the $R_s(T)$ for the Film II. The second is the breakdown of the $\omega^2$ law for the poor quality samples i.e. increasing the frequency results in an increase in the slope of $R_s(T)$.

Additional characterization of Film II was carried out in the parallel plate rig by examining the dependence of $R_s(21\ K)$ on the microwave power ($H_{rf}$) and external dc magnetic field ($H_{dc}$) as shown in Fig. 3. The data is scaled to 10 GHz. The dc and rf fields were applied parallel to each other and parallel to the film surface. Film II demonstrates an onset of nonlinearity at $H_{rf} \sim 10\ mT$. This field is close to the measured lower critical field $H_{c1}(21K)$ [20]. Of equal interest is the co-incidence of the onset of nonlinearity in dc and rf fields. Taken together this suggests that the film is not limited by weak link behavior [21], (which is consistent with dc transport observations). For reference, figure 3 also shows the performance of an all YBCO parallel plate resonator under the same conditions. The $MgB_2$ film has an $R_s$ value at low power that is about an order of magnitude greater than YBCO at the same absolute temperature and an onset of nonlinearity about an order of magnitude lower in $H_{rf}$.

**Discussion**

Previously [15] we have suggested that the temperature dependence of *lambda* in the high quality film II is indicative of the presence of an anisotropic gap $\Delta(\vec{k}) = \Delta_{ab}\cos(\theta) + \Delta_c\sin(\theta)$ where $\theta$ is the asimuth angle and $\Delta_{ab}$ and $\Delta_c$ are the values of the gap in ab-plane and c-direction of the crystal. Values of these gap are 7.2 meV and approximately 3 meV correspondingly. In a fully c-axis oriented sample current will only flow in ab-plane so that the temperature dependence of $R_s(T)$ and $\Delta\lambda(T)$ reflects the large gap $\Delta_{ab}$. In randomly oriented samples the small gap $\Delta_c$ plays the major role. The value of the large gap was extracted from the microwave measurements of the $\Delta\lambda(T)$ on Film II [15]. Further support for this scenario comes from point contact tunnelling measurements [22], which show the small gap only $\Delta_c$ = 2.5 - 2.8 meV over approximately 75% of the film surface, i.e. Film II appears to be partially c axis oriented and has a small gap value along this axis.

From the comprehensive set of experiments performed on these samples it seems that it is already possible to prepare rather high quality films. From the microwave studies of $\Delta\lambda(T)$ and the point contact studies it is already possible to see features that are intrinsic to the underlying superconductivity. However, even in the highest quality films studied so far, the absolute value of $R_{res}$ is unlikely to be close to an intrinsic value because it is a rather sensitive measure of extrinsic defects and second phase material. In addition surface roughness and MgO will both produce extrinsic $R_{res}$ losses. Comparison of film II and data taken from [14, 23] are shown in Fig. 4 and it is clear that film growth processing is far from optimized.

Nevertheless one promising route to high quality films appears to be by the preparation of an amorphous boron layer using e-beam evaporation or PLD with subsequently annealing in pure Mg vapor at 800-900°C [17, 23]. As yet it appears more challenging to produce high quality microwave films by PLD using an $MgB_2$ target, probably in part because the film quality depends on the target quality.

**Conclusion**

In conclusion, the temperature dependence of $R_s$ at different frequencies is presented. It is found that in case of the low quality samples the $\omega^2$ scaling law breaks down. In addition, in high quality films, no evidence for granularity was observed from the dc magnetic field dependence of the microwave loss. The most promising route to producing films with high quality microwave performance has been discussed. Further optimisation of film growth is needed if $MgB_2$ films are to be competitive with HTS materials for microwave applications.

**Acknowledgements**

The work is supported by the EPSRC GR/M67445, GR/R55467, NPL, Royal Society and U.S. Department of Energy under contract DE-AC05-00OR22735, etc.

# References

1 J Nagamatsu, N Nakagawa, T Muranaka *et al*, Nature **410**, 63 (2001).

2 Y Bugoslavsky, G K Perkins, X Qi *et al*, Nature **410**, 563 (2001).

3 C F Liu, S J Du, G Yan *et al.*, cond-mat/0106061.

4 X L Wang, S Soltanian, J Horvat *et al.*, cond-mat/0106148.

5 HongLi Suo, C Beneduce, M Dhalle *et al.*, cond-mat/0106341.

6 M Paranthaman, C Cantoni, H Y Zhai *et al.*, Appl. Phys. Lett. **78**, 3669 (2001) [cond-mat/0103569].

7 R S Gonelli, A Calzolari, D Daghero *et al.*, cond-mat/0105241.

8 G Burnell, D -J Kang, H N Lee *et al.*, cond-mat/0106562.

9 H Piel, Nuclear Instruments and Methods in Physics Research **A287**, 294-305, (1990).

10 A A Zhukov, L F Cohen, K Yates *et al.*, Supercond Sci. Technol. **14** L13 (2001) [cond-mat/0103587].

11 F Manzano, A Carrington, cond-mat/0106166.

12 N Hakim, P V Parimi, C Kusko *et al.*, cond-mat/0103422.

13 Yu A Nefyodov, M R Trunin, A F Shevchun *et al.*, cond-mat/0107057.

14 N Klein, B B Jin, J Schubert *et al.*, cond-mat/0107259.

15 A A Zhukov, L F Cohen, A Purnell *et al.*, cond-mat/0107240.

16 A Berenov, Z Lockman, X Qi *et al.*, cond-mat/0106278.

17 H Y Zhai, H M Christen, L Zhang *et al*, cond-mat/0103618.

18 N McN Alford *et al.*, J. of Supercond. **10**, 467 (1997)

19 L Hao, J Gallop, A Purnell, *et al.*, J. Supercond. **14**, 31 (2000).

20 A. A. Polyanskii. A. Gurevich, J. Jiang et al., submitted to Supercond. Sci. and Tech.

21 M. K. Bhide, R.M. Kadam, M.D. Sastry et al., to be published in Supercond. Sci. and Tech.

22 Y Miyoshi *et al*. Point Contact studies on MgB2 compounds - preprint

23 S Y Lee, J H Lee, J H Lee *et al*., cond-mat/0105327.

Figure Captions

**Figure 1**: Temperature dependence of the dc resistivity of films I and II.

**Figure 2**: Temperature dependence of the microwave surface resistance. All data was rescaled to 10 GHz using the $\omega^2$ scaling law. Powder – 2.5 GHz – (solid squares), data from [10]; 8.7 GHz – (open squares), Film I - 3.5 GHz – (solid circles), data from [15]; 8.7 GHz – (open circles), Film II - 3.5 GHz-(solid up triangles), data from [15]; 8.7 GHz – (open up triangle), 5 GHz – (down triangles).

**Figure 3**: Dependence of the surface resistance $R_s$ on $H_{rf}$ (solid circles) and $H_{dc}$ (open circles) for film II at 21K. For comparison, the performance of an YBCO parallel plate resonator is shown under the same conditions, where $R_s(H_{rf})$ – (solid up triangles) and $R_s(H_{dc})$ – (open – up triangles). All the data shown has been rescaled to 10GHz.

**Figure 4**: Temperature dependence of the microwave surface resistance taken from various sources [14, 15, 23]. All the data was rescaled to 10GHz using the $^2$ scaling law.

Figure 1 A.A. Zhukov *et al*.

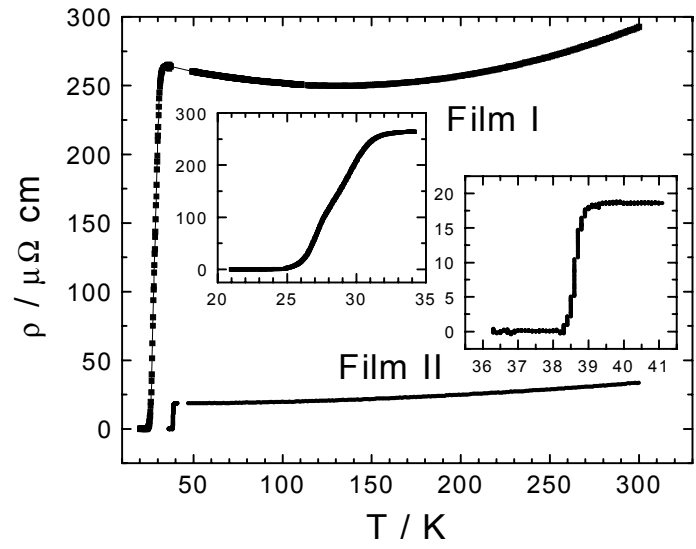

Figure 2 A.A. Zhukov *et al*.

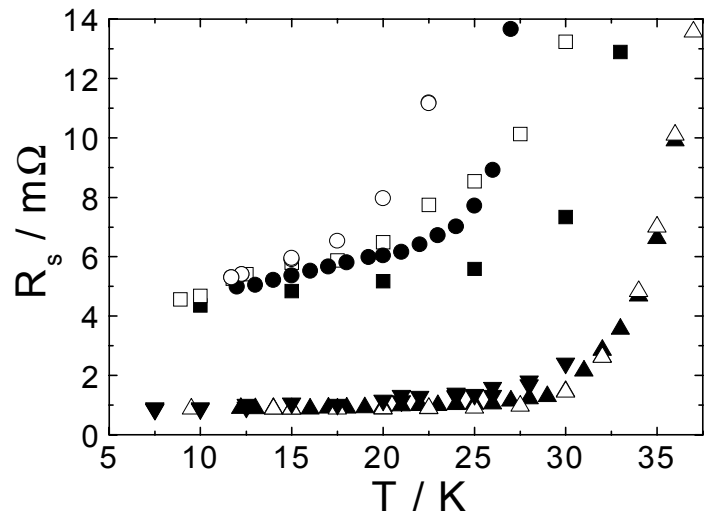

Figure 3 A.A. Zhukov *et al*.

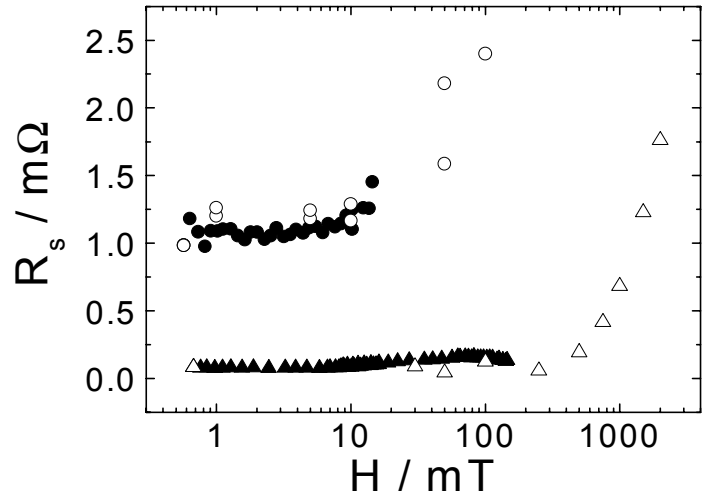

Figure 4 A.A. Zhukov *et al*.

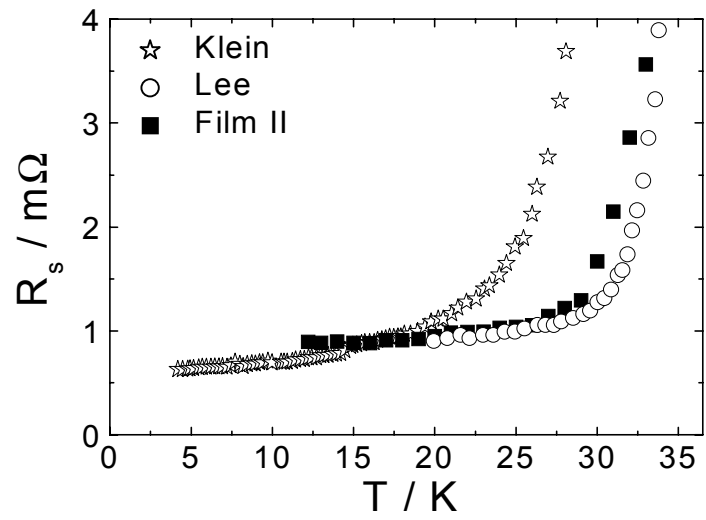